\documentclass[
reprint,
superscriptaddress,
nofootinbib,
amsmath,
amssymb,
aps,
pra,
floatfix,
longbibliography,
]{revtex4-1}

\usepackage{pdfpages}

\makeatletter
\AtBeginDocument{\let\LS@rot\@undefined}
\makeatother

\usepackage{pgffor}

\usepackage[utf8]{inputenc}
\DeclareUnicodeCharacter{00A0}{~} 

\usepackage{siunitx}
\usepackage{graphicx}

\usepackage{dcolumn}
\usepackage{bm}
\usepackage{comment}
\usepackage{chemformula}
\usepackage{hyperref}
\hypersetup{colorlinks=true,linkcolor=blue, filecolor=blue, urlcolor=blue,  citecolor=blue}

\preprint{APS/123-QED}

\begin{document}

\title{Pentagonal Photonic Crystal Mirrors:\\
Scalable Lightsails with Enhanced Acceleration via Neural Topology Optimization}

\author{L. Norder}
\affiliation{Department of Precision and Microsystems Engineering, Delft University of Technology, Mekelweg 2, 2628CD Delft, The Netherlands}

\author{S. Yin}
\affiliation{School of Engineering, Brown University, USA}

\author{M. J. de Jong}
\affiliation{Department of Precision and Microsystems Engineering, Delft University of Technology, Mekelweg 2, 2628CD Delft, The Netherlands}
\affiliation{Kavli Institute of Nanoscience, Department of Quantum Nanoscience,
Delft University of Technology, Lorentzweg 1, 2628CJ Delft, The Netherlands}

\author{F. Stallone}
\author{H. Aydogmus}
\author{P.M. Sberna}
\affiliation{Else Kooi Lab, Delft University of Technology, Feldmannweg 17, 2628 CT Delft, The Netherlands }

\author{M. A. Bessa}
\email{miguel\_bessa@brown.edu}
\affiliation{School of Engineering, Brown University, USA}

\author{R. A. Norte}
\email{r.a.norte@tudelft.nl}
\affiliation{Department of Precision and Microsystems Engineering, Delft University of Technology, Mekelweg 2, 2628CD Delft, The Netherlands}
\affiliation{Kavli Institute of Nanoscience, Department of Quantum Nanoscience,
Delft University of Technology, Lorentzweg 1, 2628CJ Delft, The Netherlands}

\date{\today}

\begin{abstract}
The Starshot Breakthrough Initiative aims to send one-gram microchip probes to Alpha Centauri within 20 years, using gram-scale lightsails propelled by laser-based radiation pressure, reaching velocities nearing a fifth of light speed. This mission requires lightsail materials that challenge the fundamentals of nanotechnology, requiring innovations in optics, material science and structural engineering. Unlike the microchip payload, which must be minimized in every dimension, such lightsails need meter-scale dimensions with nanoscale thickness and billions of nanoscale holes to enhance reflectivity and reduce mass.
Our study employs neural topology optimization, revealing a novel pentagonal lattice-based photonic crystal (PhC) reflector. The optimized designs shorten acceleration times, therefore lowering launch costs significantly. Crucially, these designs also enable lightsail material fabrication with orders-of-magnitude reduction in costs.  
We have fabricated a $60 \times 60$ \si{mm^2}, 200 \si{\nano\meter} thick, single-layer reflector perforated with over a billion nanoscale features; the highest aspect-ratio nanophotonic element to date. We achieve this with nearly 9,000 times cost reduction per m$^2$. Starshot lightsails will have several stringent requirements but will ultimately be driven by costs to build at scale. Here we highlight challenges and possible solutions in developing lightsail materials - showcasing the potential of scaling nanophotonics for cost-effective next-generation space exploration.
\end{abstract}

\maketitle

Currently, the human-made object furthest from Earth is the Voyager 1 \cite{Nasa}. Traversing space since 1977, this spacecraft has only recently left our solar system, a mere 0.5\% of the distance to the nearest star outside our solar system; Alpha Centauri. With existing propulsion systems, approaching our nearest interstellar neighbour would take over 10,000 years. In 2016, the Breakthrough Prize Foundation announced the Starshot Initiative to push the development of low-mass microchip satellites with cameras, sensors, and probes accelerated to high speeds by low-mass lightsails \cite{Starshot}. The Starshot Mission leverages advances in nanotechnology to achieve low-mass objects, and progress in high-power lasers to directionally beam energy to distant locations as far as tens of millions of kilometers away. This microchip approach to space exploration aims to reach Alpha Centauri (i.e., the nearest star outside our Solar system) within 20 years by reaching speeds up to 20\% of the speed of light, made possible by the rapidly advancing field of nanotechnology and the future potential of next-generation laser systems. Regardless of the payload mass, this mission is fundamentally exploring the physical limits of mass acceleration and our ability to reach relativistic speeds with novel mesoscopic objects made possible by nanotechnology. Of the many ambitious developments required by the Starshot Initative, the lightsails are generally considered one of the most challenging components to realize due to their unique geometries and stringent performance requirements.  

\begin{figure}
    \centering
    \includegraphics[width=1\linewidth]{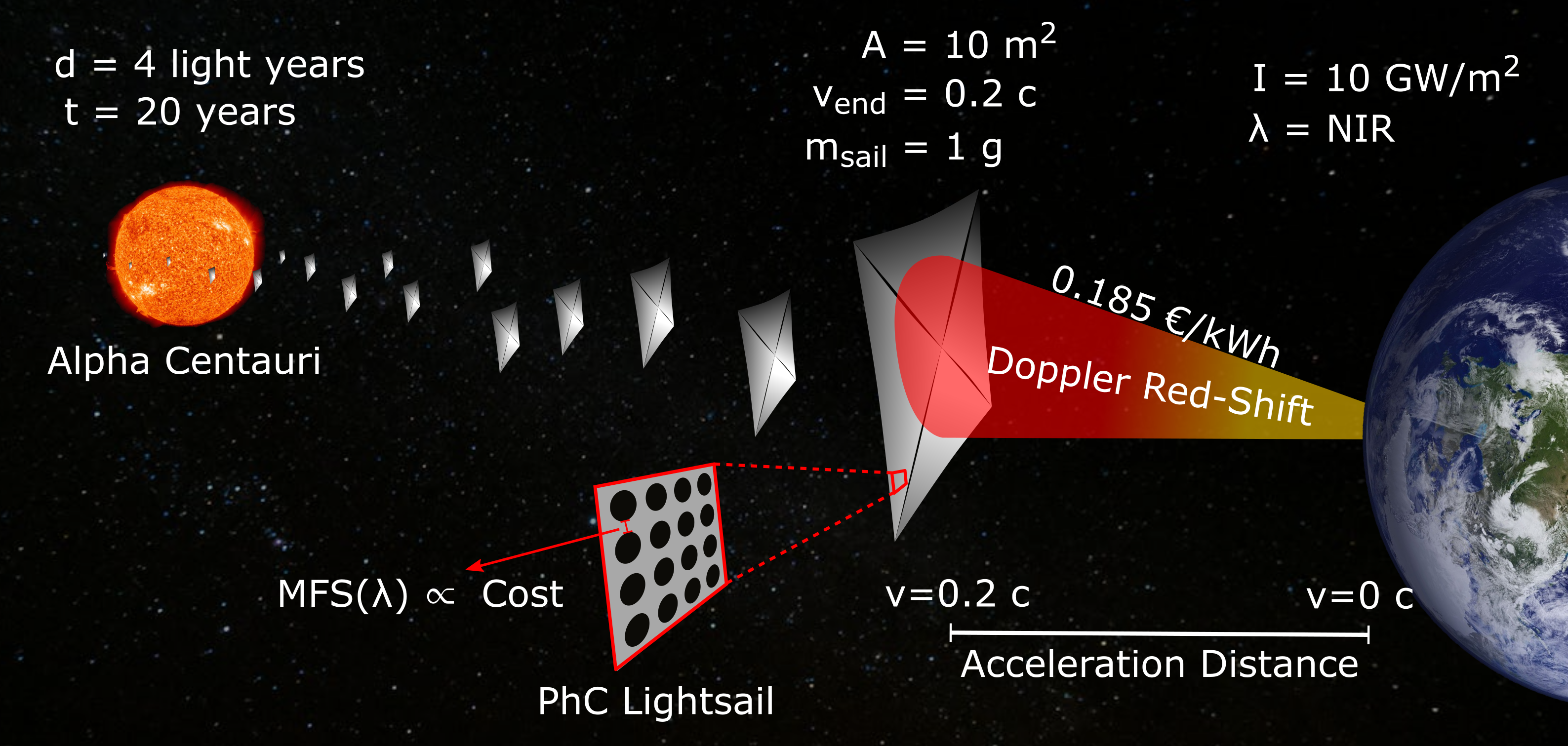}
    \caption{High power earth-based laser propelling a fleet of lightweight sails to 20\% of the speed of light, to reach Alpha Centauri in 20 years \cite{Starshot}. The lightsail needs to be reflective over a broad bandwidth due to the Doppler red-shift of the laser resulting from the change in velocity of the sail. The minimum feature size of a photonic crystal based lightsail is related to the fabrication cost. A commonly used performance metric for a lightsail is the acceleration distance. The launch cost is mainly determined by the energy consumption of the laser \cite{EUEprice}.}
    \label{fig:StarShot}
\end{figure}

The Starshot concept, presented in Fig. \ref{fig:StarShot}, is based on generating an optical force on a reflective lightweight sail material by projecting a high-power Earth-based laser on it. As proposed in the Starshot initiative, the lightsail will be approximately 10 \si{m^2} and the laser power 10-100 \si{\giga\watt\per\meter^2} to generate sufficient radiation pressure \cite{Starshot} within a few minutes of laser exposure. To approach relativistic speeds (0.2c), stringent low-mass budgets are required, limiting the weight of the sail and the connected payload chip to approximately 1~gram each. The laser used for radiation pressure on the sail needs to operate on wavelengths in the near-infrared (NIR) spectrum from 1 - 2 \si{\micro\meter} because of its low atmospheric absorption \cite{hemmati2013}. These lightsails will experience Doppler-shifts as they accelerate, requiring high broadband reflectivity \cite{Atwater2018}. Larger bandwidth can generally be achieved by increasing the thickness of the sails at the cost of additional mass, which can severely reduce its acceleration performance. Given the interaction with a high-power laser beam, lightsails must achieve ultra-low optical absorption to avoid thermal fracturing. While all components from payload to lightsails will require significant development over the next decades, the lightsail stands out as the major challenge of this initiative because of its unique requirements. Achieving a 1~g microchip payload will require miniaturizing all of its components like cameras, communications, and sensors in x, y and z dimensions. On the contrary, achieving gram-scale, 10 \si{m^2} lightsails will require spanning a reflector to meter scales in x and y while retaining nanoscale thickness -- far from any aspect-ratio achievable by modern nanotechnology. The physics and economics of how these high-aspect ratio reflectors are made will be crucial to the success of this technology.   

One often neglected aspect of this mission is that these long-distance missions rely on a shotgun approach of many sails to increase the chance of success. This means the costs of manufacturing and launching these sails with high power (for several minutes) are major considerations that have not been taken into account in the design process of the sails but are crucial to Starshot's ambitious goals. 

Many possible lightsail materials are proposed in the literature \cite{Jin2020, Myilswamy2020, Gao2024, Atwater2018, Moura2018, Jaffe2023, Gao2022}. Among these materials, single-layer silicon nitride (SiN) photonic crystals are the top candidate material because SiN combines low optical absorption and the low mass and high reflectivity achieved by single-layer hole-based photonic crystals. Advantageously, silicon nitride is a well-studied and mature CMOS material that can be conventionally integrated with many microchip platforms. Photonic crystals made from SiN have been well studied in the field of optomechanics, which also favours low mass and absorption with high reflectivity \cite{stambaugh2015, Makles2015, Land2024, Norte2016, Chen2017,bernard2016precision, enzian2023phononically}. Additionally, \ch{SiN} membranes will not wrinkle due to the internal tensile stresses generated in the deposition allowing for better stability once suspended. This pre-stress in SiN photonic crystals allows for precise alignment of optical beams onto the suspended photonic crystals in a lab-scale test setup. Due to these favourable properties, \ch{SiN} is chosen as the lightsail material for this work.

Given that photonic crystal reflectors rely on a two-dimensional array of subwavelength holes in a single-layer SiN membrane, it is important to note that there is a direct relation between the minimum feature size (MFS) of the patterns (e.g., minimum distance between holes), and the costs of manufacturing the lightsail; lower MFS means higher costs (more intricate geometric details) but potentially lower mass and better acceleration. This sets up a complex trade-off between cost \cite{lubin2022economics}, manufacturing and acceleration performance that has not been previously considered. Additionally, a bigger MFS and larger surface area of the sail, can be favourable to crucial properties like stress reduction and increased radiative cooling.

Although single-layer photonic crystals have proven to be effective reflectors even with simple two-dimensional hole lattice designs, the few contributions targeting lightsail design have not considered state-of-the-art manufacturing constraints. The traditional optimization of photonic structures is highly iterative and relies on domain knowledge from experienced researchers \cite{Liu2018}. This trial-and-error process is unlikely to be successful in finding high-performance designs because of the high-dimensional design space. Additionally, photonics optimization is usually non-convex, resulting in a challenging optimization. Notwithstanding, inverse design methods have resulted in promising, non-trivial and high-performance PhC designs \cite{Jensen2011, Li2019, Molesky2018, Campbell2019} even for lightsail design \cite{Jin2020}.

Recently, a new inverse-design method referred to as neural topology optimization (neural TO) has been proposed where conventional TO is enhanced by machine learning via the reparameterization approach proposed by Hoyer et al. \cite{hoyer2019neural}. This strategy differs from most machine learning contributions aimed at improving inverse design methods. Usually, machine learning is used in inverse design by training generative models such as variational autoencoders and generative adversarial neural networks  \cite{Ma2018, Jiang2019a, Peurifoy2018} that require large training databases and have difficulties with predictions that fall out of the training data distribution. In contrast, neural TO introduces a neural network before a physics solver (e.g. finite element analyses) and shifts the optimization problem to finding the weights and biases of the neural network that minimize the objective function calculated by the physics solver. Neural TO is still in its infancy and has not been applied in the context of inverse problems in photonics. However, in this work we find that the method is particularly advantageous for lightsail design when compared to conventional TO strategies. Additional information regarding the employed optimization algorithm is presented in the supplementary information.

The primary objective of this study is to design a PhC lightsail that maximizes acceleration capabilities while minimizing mission costs by addressing both the MFS constraints imposed by lithography processing and the costs associated with laser time needed to accelerate the lightsail. Optimizing solely for acceleration capabilities can lead to designs with a small Area fraction (less mass)~\cite{Jin2020} which are more delicate and difficult to fabricate and launch. Conversely, the costs of manufacturing high-yield lightsails and producing minimum feature sizes tend to result in designs with higher Area fraction. To navigate this complex parameter space, neural topology optimization is adapted to meet these lightsail design challenges. We then show we can produce these wafer-scale lightsail materials at nearly three orders-of-magnitude reduction in costs. 
\section*{Results}

\textbf{Photonic Crystal Lightsail Design.} The stringent Starshot mission requirements have driven research on free-standing photonic crystals (PhCs) as broadband reflectors. PhCs control light propagation by tuning sub-wavelength variations in refractive index materials \cite{Joannopoulos2008}. 

\begin{figure}
    \centering
    \includegraphics[width=1\linewidth]{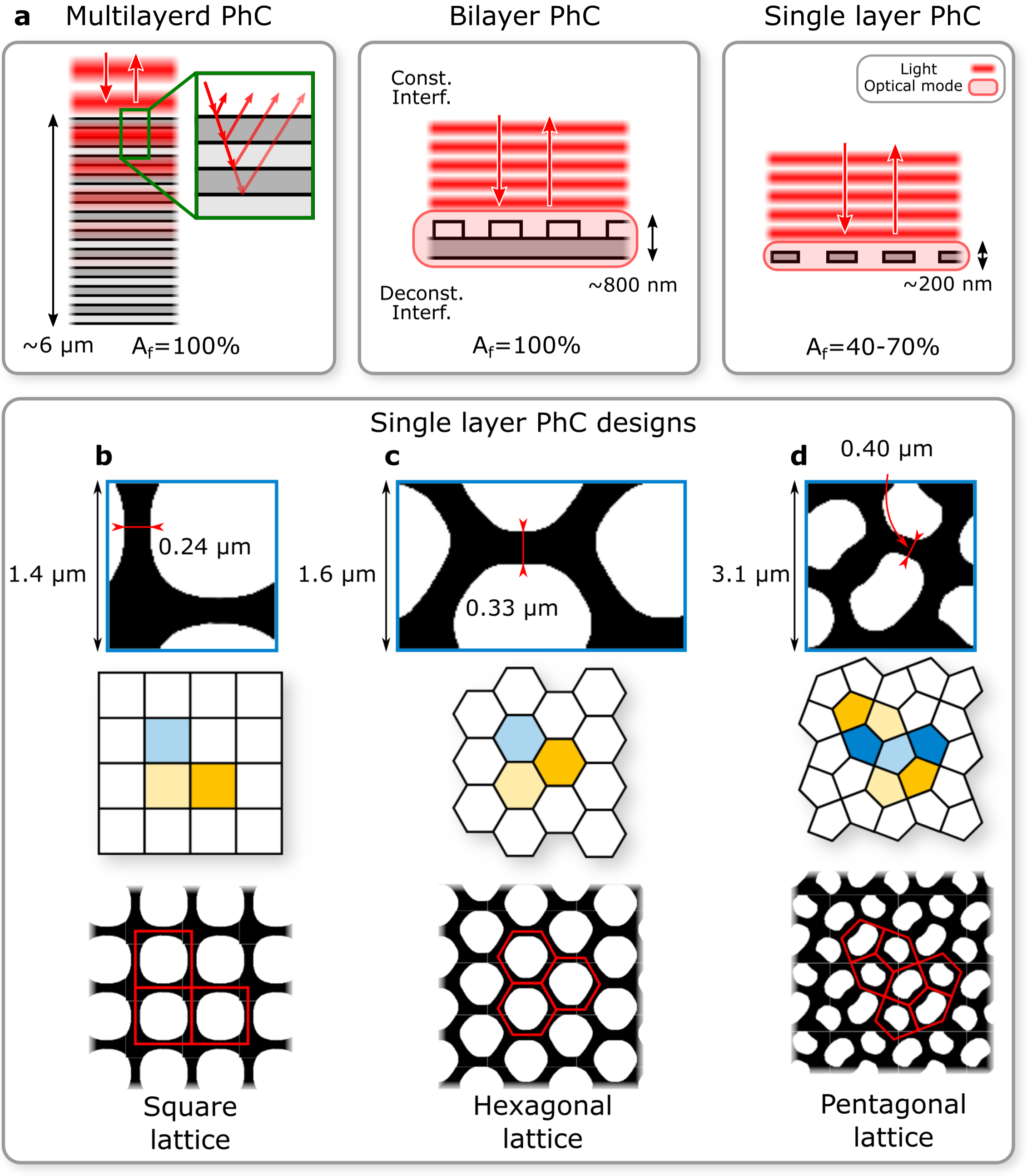}
    \caption{\textbf{a}, Working principles of different photonic crystal architectures. Multilayered PhC consists of stacked layers with varying refractive indices. The bilayer PhC consists of a repeating PhC pattern on top of a solid membrane. Single layer PhC is a membrane with a repetitive PhC hole pattern. For both the bilayer and single-layer PhC, the incident light creates an optical mode within the material that deconstructively interferes with the transmitted light and constructively with the reflected light. The best optimized single layer PhC design without area constraint for square lattice (\textbf{b}) and hexagonal lattice (\textbf{c}), where black is material and white is vacuum. The square and hexagonal lattice thicknesses are 0.2 $\mu m$ and 0.3 $\mu m$ respectively. \textbf{d}, The pentagonal lattice design for an Area fraction $A_f$ of 55\% with a thickness of 0.18 $\mu m$.}
    \label{fig:des}
\end{figure}

Figure \ref{fig:des}a illustrates the working principles of different PhC architectures. The most well-known reflectors for mirror coatings are multilayered photonic crystals, or distributed Bragg reflectors (DBRs), which consist of several layers of dielectric materials with alternating refractive indices and subwavelength thicknesses. These multilayered PhCs, typically several microns thick, can achieve very high reflectivity ($>99.5$\%) over a broad bandwidth ($\approx$200 nm). However, they are too massive for Starshot's requirements (hundreds of grams for a 10~$m^2$ sail), and their ultra-high reflectivity is not particularly useful for acceleration.
\\
In contrast, single-layer photonic crystals achieve changes in refractive index through periodic holes in a membrane, providing alternating refractive indices in the x and y directions. Incoming light creates an optical mode in the membrane that constructively interferes with incoming light and destructively interferes with transmitted light, resulting in high reflectivity ($\approx$~98.9\%) over a narrower range (20 nm). This design offers an ultra-thin geometry. Due to their design flexibility and single-layer nature, two-dimensional PhCs are expected to offer higher reflectivity for a lower mass \cite{Moura2018}, as the small film thickness and holes reduce nearly half the mass. Single-layer PhCs are currently the only architecture thin enough to achieve a 1-gram lightsail. However, evaluating their acceleration properties rigorously requires considering broadband reflection, mass, and laser divergence.
\\
Bilayer photonic crystals offer a hybrid approach by increasing thickness (up to a micron) \cite{Jaffe2023, Gao2024, Chang2024, Brewer2022, salary2020photonic} to enhance reflection bandwidth (Fig. \ref{fig:des}a), but this trade-off adds mass, significantly impacting the sail's acceleration. Recent efforts have explored a bilayer photonic crystals \cite{Chang2024}, which consist of one uniform layer and an additional layer of single-layer photonic crystal on top to increase the bandwidth and ease fabrication. However, these types of reflectors can quickly perform worse in terms of acceleration than a single-layer, unpatterned (low-reflectivity) SiN membrane, which serves as our standard worst-case scenario. For instance, bilayer PhCs with a SiN PhC on a Si layer with an average reflectivity of 80\% have a similar acceleration distance of 67 Gm compared to a 35\% reflective and 200 nm thick unpatterned SiN membrane, as shown in Fig. 3a (red line). This is in agreement with Atwater et. al (2018) \cite{Atwater2018}, and underscores the challenging mass requirements for designing lightsails thicker than a single layer to achieve broader bandwidths. Consequently, given the stringent mass requirements, we focus on single-layer PhCs, which are more suitable for lightsail design than other PhC architectures. The supplementary information includes a study discussing the limits of the initially proposed mass target using two-dimensional PhC sails.

\textbf{Acceleration Performance vs. Costs.} In designing the lightsail, we must consider not only its acceleration performance but also the associated costs, including those resulting from lithography, manufacturability, and yield, which ultimately impact the final costs. This complexity arises because optimizing for acceleration often leads to designs with low Area fraction and thickness, while manufacturing costs and yield would be significantly reduced with high Area fraction. This sets up a challenging set of tradeoffs in designing and optimizing a photonic crystal that balances both acceleration performance and costs.
\\
In terms of cost, we focus on the lithography process to reduce the fabrication cost as it takes the most time and money compared to other fabrication steps, especially when scaled to square meter-sized PhCs. Selecting a fitting lithography method for patterning the PhC-based lightsail is an integral part of the nano-photonics fabrication due to its direct impact on the achievable resolution and writing speed. E-beam lithography is commonly used for nm-sized structures, yet it is slow and expensive for more extensive areas \cite{Chen2015}. E-beam writing times for 1 \si{cm^2} can vary from multiple days for conventional techniques~\cite{Moura2018} to numerous hours for the most advanced methods \cite{Trasobares2014, Chang2024}. However, a faster and more affordable nanofabrication method is optical lithography \cite{Sharma2022, Brink2019}. Therefore, i-line photolithography (i.e. light source with 365 nm light), which has an MFS of 500 \si{nm}, was selected for this study based on its cost-effectiveness, availability, and established processing protocols. Additionally, the writing time is independent of the design because of the use of a mask, making it a good match for the possible irregular and non-trivial design generated by the neural TO. In the supplementary information, a more detailed comparison is made between the different photolithography methods.
\\
The writing costs are related to the operating costs of a cleanroom, which are expected to be around 200 euro/hr. When choosing optical over e-beam lithography, the writing time of a 10 \si{\meter^2} sail can markedly be reduced from 15 years to one day, calculated with $7.5\times 10^{-5}$ \si{m^2/hr} and 0.43 \si{m^2/hr} respectively. Therefore, the cost can be reduced almost 9000 times, from 26 million euros to 3,000 euros per sail.
\\
Within the Starshot initiative, there is no agreement yet on which wavelength the laser uses. Specifically for the lightsail development, 1550 nm is the preferred wavelength because the feature size of the PhC is proportional to the wavelength. Therefore, the fabrication cost and complexity are reduced due to the larger features. Furthermore, the optical absorption of SiN is lower for 1550 nm light \cite{Steinlechner2017}, allowing for the use of high-power lasers. As an additional benefit, the atmospheric absorption of 1550 nm light is less than other wavelengths in the NIR \cite{hemmati2013}. 
\\
The Starshot mission not only emphasizes reducing mass through nanotechnology but also harnesses advancements in arrayed lasers to project energy directionally across vast distances, optimizing propulsion efficiency. The high-power lasers for the Starshot mission are expected to operate at a single wavelength. As the sail accelerates to high speeds, Doppler red-shifting will alter the wavelength of the light relative to the sail. This necessitates that these ultra-thin reflectors remain highly reflective over the Doppler bandshift. However, there is an inverse relationship between the thickness of the reflectors and their reflectivity bandwidth: ultra-thin reflectors exhibit high reflectance over a narrow band, while thicker reflectors, which can increase the bandwidth, also add significant mass. This added mass can hinder acceleration. Thus, balancing thickness and broadband operation is a major challenge, which is evaluated using a figure of merit that includes reflectivity, mass, and the Doppler shift of the sail. 
\\
Initially, it is chosen to minimize the acceleration distance ($D$), i.e. the distance required to reach the final velocity of the lightsail, as the optimization objective. This quantity of interest is commonly used in lightsail design as it enforces a tradeoff between weight and broadband reflectivity \cite{Macchi2009}. Furthermore, it implicitly takes into account the laser's divergence limits \cite{Kulkarni2018}. The MFS imposed by the fabrication method must be included in the optimization to consider the fabrication cost. However, controlling the MFS explicitly is a challenging problem in the TO field \cite{Sigmund2013} and becomes even more non-trivial for a neural network-based TO. Therefore, the MFS is generally controlled implicitly \cite{Zhang2016}. We extended the optimization with a simple approach of adding an Area fraction ($A_f$) as an extra optimization constraint \cite{hoyer2019neural} to control the MFS of the final design.
\\
For the final mission, the laser is presumed to emit a linear polarised plane wave. Optimizing a PhC lightsail for only one polarization direction results in parallel strings aligned with the polarisation \cite{Jin2020}. Thus, a precise and challenging alignment of the physical sail with the laser beam position and its polarization is required. Additionally, string-based PhCs are not practical for lightsail fabrication as they would stick together. Therefore, in this study, the sail is optimized for two orthogonal polarisation angles $\phi=0$ and $\phi=\frac{1}{2}\pi$ (i.e. rotation around the normal of the crystal plane) to obtain producible two-dimensional designs and promote a polarisation direction invariant design, relaxing the alignment requirements. Additional information regarding the optimization and its formulation is described in the methods section.

\begin{figure}
    \centering
    \includegraphics[width=\linewidth]{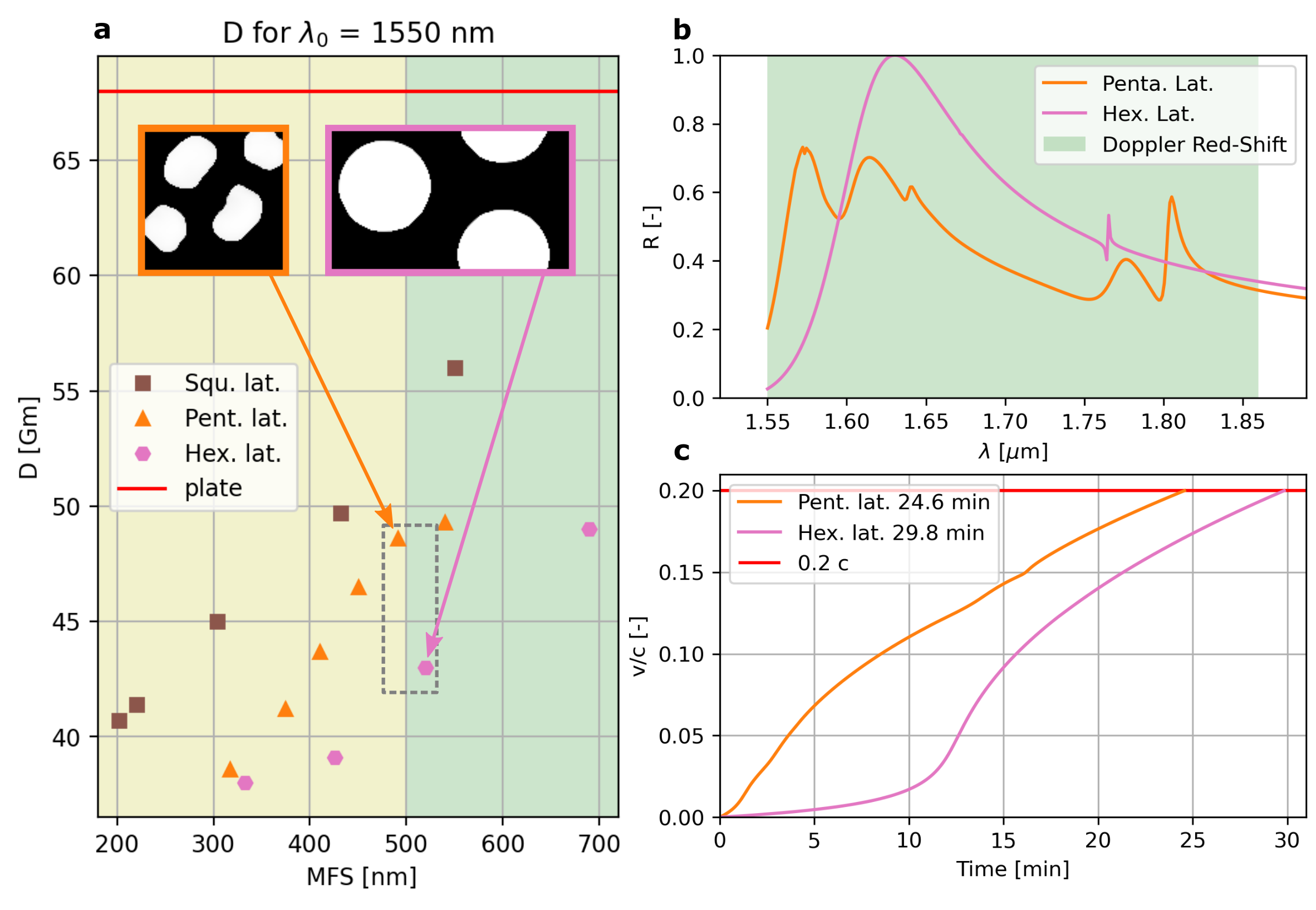}
    \caption{\textbf{a}, Acceleration distance ($D$) for different lattice structures with varying minimum features size (MFS). The red line indicates the D for a 200 nm thick un-patterned PhC membrane. \textbf{b}, The reflectivity spectrum of the selected photonic crystal designs from \textbf{a} for the full Doppler shift region. The rest of the energy is transmitted due to the ppm absorption of SiN \cite{Steinlechner2017}  \textbf{c}, The velocity of the hexagonal and pentagonal PhC lightsail during acceleration compared with the speed of light.} 
    \label{fig:MFS}
\end{figure}

\textbf{Computational Results.} 
At first, the optimization was conducted without an area constraint ($A_f$), yielding designs with patterns following conventional square and hexagonal crystal lattices (Fig. \ref{fig:des}b,c). However, these fall short of the MFS $> 500$ nm objective. Therefore, optimizations with a $A_f$ constraint from 40\% - 70\% were subsequently realized. Notably, larger $A_f$ led to a unique pentagonal lattice structure \cite{Moulton2001} (Fig. \ref{fig:des}d). The performance of various designs is compared in Fig. \ref{fig:MFS}a. The figure shows that an increasing $A_f$ correlates to an increase of MFS and, consequently, a decrease in performance (i.e. an increase of $D$).
\\
The equation of motion of the lightsail \cite{Kulkarni2018} is solved directly from the reflectivity spectrum presented in Fig. \ref{fig:MFS}b to obtain the sail velocity over its acceleration time (Fig. \ref{fig:MFS}b), so the relation between the reflectivity spectrum and the performance can be studied. Intuitively, one would think that a design with a larger acceleration distance will also take more time to be accelerated. However, the key insight obtained from Fig \ref{fig:MFS}c is that this is not the case. The pentagonal design with a higher D than the hexagonal design has a significantly lower acceleration time.

\begin{figure}
    \centering
    \includegraphics[width=\linewidth]{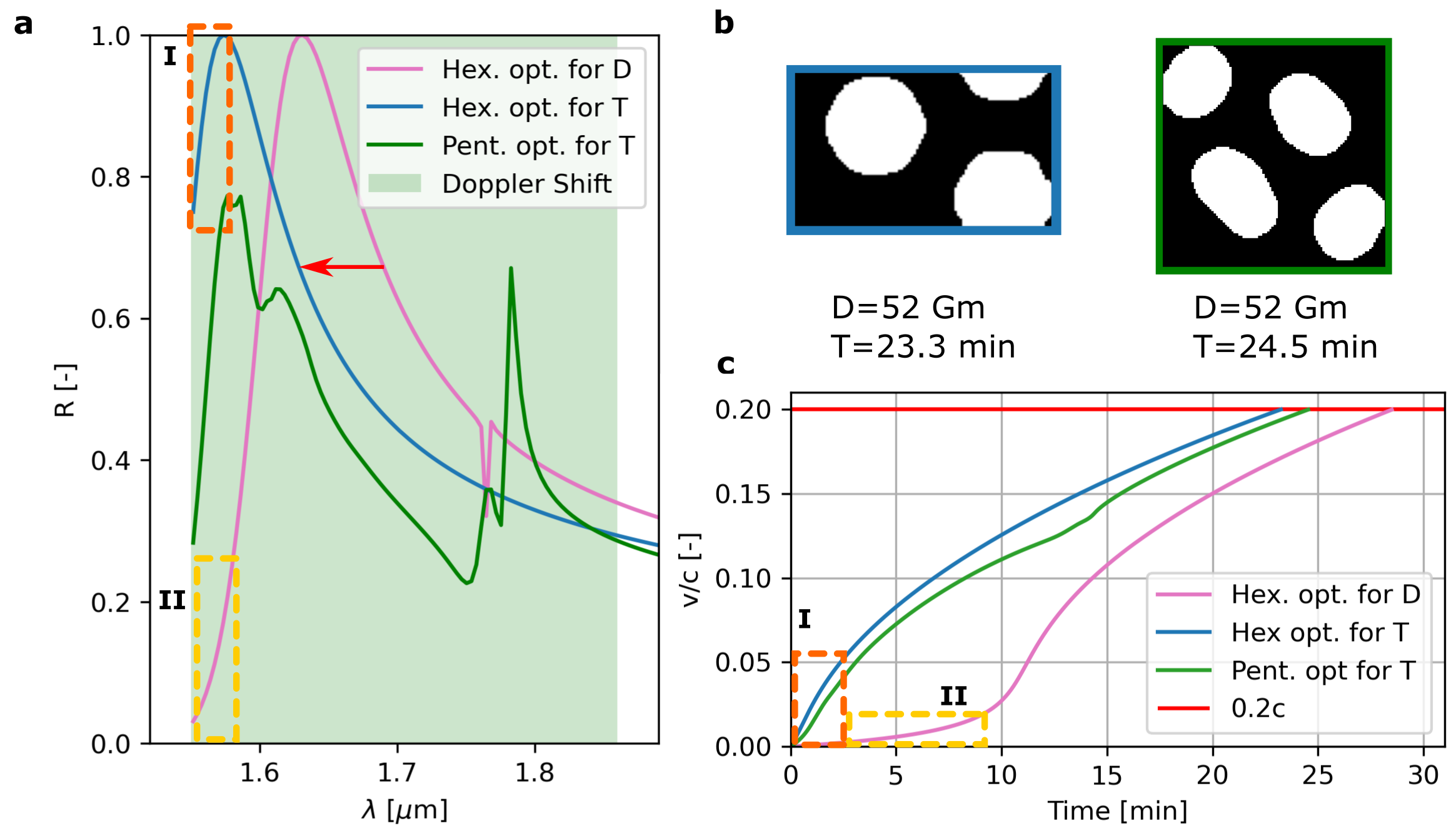}
    \caption{\textbf{a}, Reflectivity spectrum of two hexagonal PhCs. The red arrow indicates the shift of the reflectivity peak to the laser wavelength (i.e. to the left) when optimizing for T.\textbf{b}, the final design of hexagonal (blue) and pentagonal (green) PhC optimized for T. \textbf{c}, The velocity of the PhC lightsail during acceleration compared with the speed of light. Regions I (orange) and II (yellow) indicate how the beginning of the reflectivity spectrum translates to the initial acceleration of the sail.} 
    \label{fig:DvsT}
\end{figure}

 Additionally, the pentagonal design obtained by the neural TO method gives the non-trivial insight that a broadband reflector can be made with a two-dimensional PhC by designing it with multiple hole sizes and shapes, resulting in multiple resonance peaks. However, these peaks' total reflectivity is lower than that of a PhC designed with one fixed shape, thereby making the reflector more broad-band rather more than reflective. This allows a sail to be tuned to more wavelengths within the Doppler range, a quality not usually needed for conventional mirrors but critically important for lightsails. The supplementary information provides a more comprehensive analysis of the obtained designs, including the polarization dependence, acceleration distance, and time.
\\
Notably, the launch cost is only determined by the acceleration time (T) (i.e. the time required to reach the final velocity of the lightsail), making it a significant performance parameter to consider. For example, when assuming ideal energy conversion to the laser and ideal momentum transfer to the sail, the time difference of 5 minutes between the pentagonal and hexagonal lattice, shown in Fig. \ref{fig:MFS}c, can mean a difference in launch cost of 1.5 million euros with respect to a total launch cost of 9.3 million euros when calculating for a 10 \si{GW/m^2} laser on a 10 \si{m^2} sail and 0.185 euro/kWh \cite{EUEprice} (average non-household energy price 2023). Considering the high throughput of launches required for Starshot missions, the costs of individual launches become of utmost importance. 
\\
Secondly, the lightsail is optimized to minimize T as the FOM because of the large impact of T on the launch cost. The formulation of the FOM can be found in the methods section. For this optimization, the obtained designs are the same as when optimizing for D. The best design from this optimization, which satisfies the MFS objective, is presented in Fig. \ref{fig:DvsT}b. Notably, the best design optimized for T follows a hexagonal lattice and has reduced the launch time by 6 minutes compared to the design optimized for D. This decrease in launch time results in a cost reduction of approximately 2 million euros per launch (i.e. following the same calculation presented above). 
\\
However, this decrease in T comes with the cost of an increased D. Fig. \ref{fig:DvsT}a,c shows that the reflectivity of the PhC for the initial wavelengths is responsible for the fast initial acceleration of the sail and, therefore, reaching its final velocity in less time. Alternatively, for a lightsail with low reflectivity at the initial wavelengths, the sail will only be accelerated slowly, wasting a lot of illumination time before it gets significantly accelerated. So, when optimizing a PhC for only $D$, initial reflectivity is not prioritized and can result in a long acceleration time. 
\\
Regarding the pentagonal lattice, optimizing for T or D did not change the performance significantly. However, when comparing the pentagonal and hexagonal designs optimized for T in Fig.\ref{fig:DvsT}b,  it can be seen that both designs follow a similar path when accelerated, meaning that the designs are close together within the design space. This can indicate the neural TO finds the final solution in a basin where different designs have comparable outcomes for the FOM. Additional mission requirements can be included in the optimisation to resolve this basin. Different requirements will call for other inherent properties of a PhC and determine the most suitable crystal structure for the lightsail application. For example, a notable difference between the pentagonal and hexagonal designs is that the $A_f$ for the pentagonal design is higher. This property can be beneficial to aid radiative cooling and reduce stress concentration within the sail during its dynamic operation. In terms of fabrication, a larger $A_f$ means more material between holes, and more robust structures. In contrast, small $A_f$ would mean PhC designs characterized by small delicate wires of materials between holes which must survive the fabrication process of suspending the structures and subsequently undergo fast accelerations. The Area fraction will be a crucial parameter affecting several other important factors, including costs from minimum feature size and manufacturability (i.e. the ability not to fracture too easily) and acceleration capabilities.     

\begin{figure}
    \centering
    \includegraphics[width=1\linewidth]{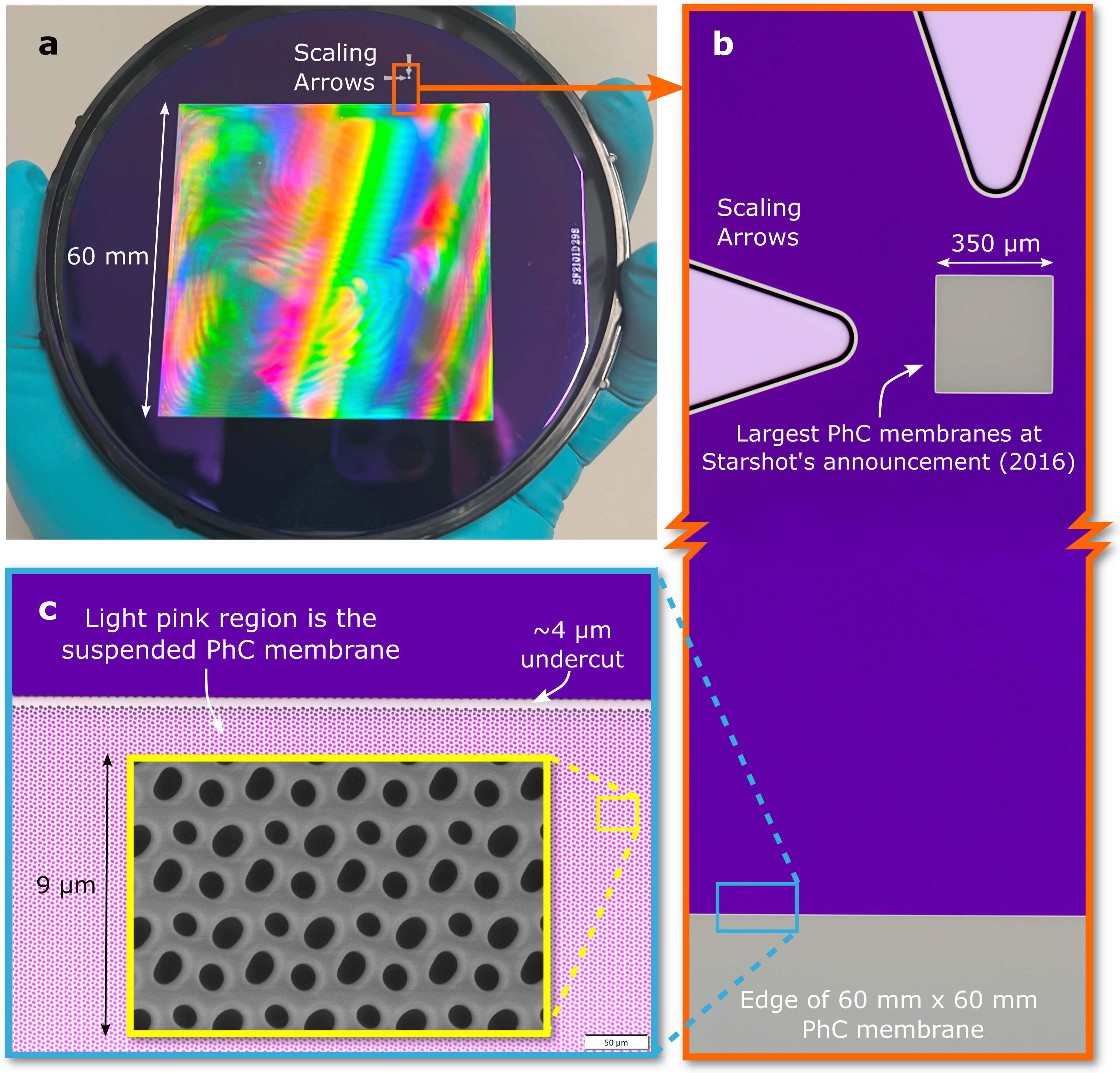}
    \caption{\textbf{a}, Photograph of a 100 mm wafer with a $60 \times 60$ \si{mm^2}, 200 nm thick suspended SiN PhC membrane, covered with a pentagonal pattern having a period of 3.0 \si{\micro\meter}.
    \textbf{b}, Microscope image of two arrows etched into the substrate pointing towards a $350 \times 350$ \si{\micro\meter\squared} suspended PhC membrane. The bottom of the orange-framed inset shows the edge of the $60 \times 60$ \si{mm^2} suspended membrane in the same magnification. The $350 \times 350$ \si{\micro\meter\squared} membrane puts the large membrane in perspective by showcasing the largest single-layer suspended PhC membranes at Starshot's announcement (2016) \cite{Chen2017}
    \textbf{c}, 50x magnification of the edge of the membrane. One can see the repeating pattern covering the $60 \times 60$ \si{mm^2} phononic crystal (PhC). The SiN is still attached to the silicon frame in the purple regions. The light pink indicates where the silicon has been removed under the PhC, leaving a suspended SiN PhC membrane. The yellow-framed inset shows a further zoom of the pentagonal lattice taken with a scanning electron microscope. }
    \label{fig:SEM-PhC}
\end{figure}

\begin{figure}
    \centering
    \includegraphics[width=1\linewidth]{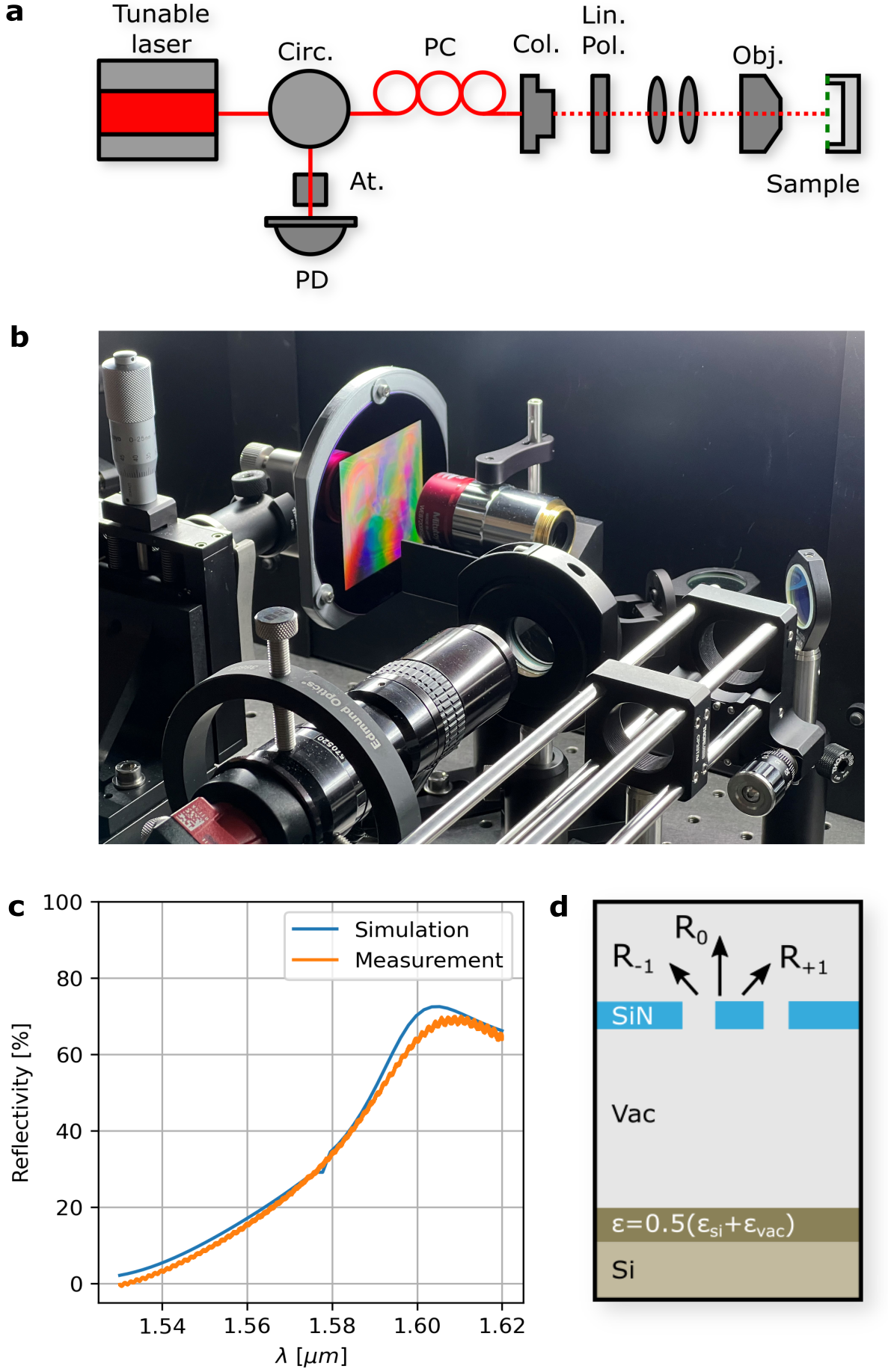}
    \caption{\textbf{a}, Experimental setup. Circ, circulator; PC, polarization controller; Col., collimator; Lin. Pol., linear polarizer; obj., objective; PD, photodetector. \textbf{b}, 100 mm diameter wafer with $60 \times 60$ \si{mm^2} suspended PhC membrane clamped in the measurement setup. \textbf{c}, Simulations from the actual fabricated design obtained from the scanning electron microscope (blue), measurement (orange) with respect to a silver mirror reference. \textbf{d}, Schematic representation of the simulation. A small layer with a relative permittivity of $(\epsilon_{Si}+\epsilon_{vac})/2$ \cite{Fujiwara2018} is introduced to represent the rough surface resulting from the undercut of the SiN membrane.}
    \label{fig:PHC_Meas}
\end{figure}

Transitioning to the broader context of existing literature, different two-dimensional PhC designs have been proposed before, with designs having a MFS between 125 - 260 nm, and $D$ between 1.9 - 13 Gm  \cite{Jin2022, Atwater2018, Brewer2022, Kudyshev2022, Ilic2018}. However, meaningful comparisons with our findings pose a challenge due to the different mission parameters employed in previous studies. Factors such as variations in payload mass, laser power, and sail material limit the direct applicability of our optimized designs to those reported in the literature. Hence, newly proposed designs in this field should ideally utilize the same mission parameters. Nevertheless, it is worth noting that, despite lower D, the designs proposed in the literature are challenging or expensive to fabricate due to their intricate features and material choices. Additionally, there is a challenging tradeoff between preferable physical properties like broadband reflectivity, stability or cooling and the thickness of the PhC \cite{Chang2024, salary2020photonic, Jin2020, Gao2024}. However, the extra mass can significantly increase the acceleration distance and time and thereby the cost.
\\
\textbf{Experimental Results.} For the reasons invoked previously, the pentagonal design presented in Fig. \ref{fig:MFS}, was chosen to be fabricated in this work as a proof of concept. Fabricating a pentagonal lattice PhC membrane also illustrates the robustness of the fabrication method (elaborated in the Methods section).
Fig. \ref{fig:SEM-PhC} shows $60\times60$ \si{mm^2} and a $350\times350 $ \si{\micro\meter\squared} suspended single-layer PhC membrane. To illustrate the large scales of these suspended devices, we have etched millimeter-scale arrows pointing towards the smaller membrane. The smaller membrane represents the largest photonic crystals made at Starshot's announcement in 2016 \cite{Chen2017}, highlighting a nearly 30,000-fold increase in surface area of SiN photonic crystal materials. Here, we are able to produce this reflective material at nearly 9000 times reduced cost per square meter. This underscores the progress in the scalability and aspect ratio achievable with our fabrication method and design methodologies that consider yield. Remarkably, the device shown in Fig. \ref{fig:SEM-PhC} is one of the largest single-layer suspended PhC to date, having the highest aspect ratio (length/thickness) of 3$\times 10^{5}$ of any nanophotonic element and covered in about 1.5 billion nanoscale holes. To give an intuitive sense of this aspect-ratio, our 200nm-thick photonic crystal scaled up to a 1mm thick glass sheet would extend for nearly $\frac{1}{3}$ km laterally, covered in $\approx$2.5mm-diameter holes with $\approx$2.5 mm of glass between holes -- an aspect ratio that is far beyond anything manufactured at macroscopic scales. At nanoscales where weight and forces scale differently due to low masses and small surface areas, unique high-aspect-ratio geometries can be reliably produced.
\\ 
\\
A tunable laser (range: 1530 - 1620 nm) is used in the measurement setup (Fig. \ref{fig:PHC_Meas}a and Methods) for obtaining a part of the reflectivity spectrum to validate the simulations. The measurement and the simulations are shown in Fig. \ref{fig:PHC_Meas}c. The measured value differs from the original spectrum due to the fabrication steps, like etching, which etch away some of the membrane's thickness during the undercut and enlarges the holes due to non-perfect selectivity. Therefore, the final shape of the PhC is retrieved via an electron microscope and used to obtain the expected reflectivity. Notably, the measurement performed is in good agreement with the simulation of the fabricated PhC. 
\\
Analysis of the final membranes revealed that the hole size of the PhC at the edge is approximately three percent larger than that of the center holes. This size difference causes a small shift in the reflectivity spectrum of 10-20 nm, likely because the etch rate in the middle is lower due to more etchant chemicals being available at the edge. Therefore, the middle, having more exposed silicon than the edges, consumes more SF$_6$ chemical (used to undercut our PhCs) and reduces the etch rate compared to the edge, which is adjacent to the substrate without holes and does not consume the chemical. This results in the membrane releasing first from the edges and then from the center, which is advantageous since we use cryogenic temperatures to improve the SiN/Si selectivity of the SF$_6$ etchant. Thus, the center remains well anchored thermally to the substrate during the release. The fabrication process can be optimized to counter the above-mentioned variations for an even better match with the optimized design. However, the suspended PhC membrane measurement is in reasonable agreement with the simulations.
\\
\section*{Discussion \& Conclusion}
High-aspect-ratio PhC reflectors, with subwavelength thickness and centimeter-scale dimensions, offer unique capabilities not achievable at smaller micron scales, as shown in Fig 5b. Centimeter-scale photonic crystals can achieve higher reflectivity with thinner geometries because they do not require light to be focused down, which can severely reduce reflectivity from the ideal case of a plane wave incident on a PhC. In these larger-scale PhCs, the incident beam can interact with billions of nanoholes, similar to a plane wave on an infinitely sized PhC \cite{Moura2018}. Their novel geometries open new possibilities for lightweight, compliant reflectors in dynamic applications like movable mirrors \cite{Madec2012}, imaging optics \cite{Xu2024b}, as well as for acceleration to high speeds in space exploration.

This study presents the fabrication of the largest single-layer suspended photonic crystal (PhC) with the highest aspect ratio achieved for a nanophotonic element, marking an advancement for large-scale PhC lightsails required for missions like the Starshot Initiative. Notably, we have achieved a 9000 times reduction in manufacturing costs, a critical breakthrough for the project's viability. This cost reduction stems from surpassing the minimum feature size (MFS) threshold set by diffraction, allowing the use of high-throughput photolithography for large wafer-scale production at significantly lower costs. We use the Area fraction of the photonic crystal as a way of optimizing for MFS, which is traditionally difficult in topology optimization. 

Previous research has focused on optimizing acceleration performance, but this study directly addresses the critical costs of manufacturing, yield (i.e., lightsail breakage), and laser launching. The Starshot project's shotgun approach highlights that economic considerations are as crucial as scientific performance for mission success. The coupling of economics and performance will ultimately determine feasibility and can lead to non-intuitive design strategies. 

The design process was conducted by neural topology optimization (neural TO). This method was found to be more robust than performing topology optimization without the neural network reparameterization trick, as it does not require artificial relaxation in the simulations to converge to optimum solutions. Traditional PhCs can be highly reflective at a single wavelength peak or more broadband by increasing thickness, which adds mass. Constrained to a single layer of SiN, neural TO discovered a basin of possible designs with similar performance and a novel periodic placement of holes: the pentagonal lattice. This lattice features several peaks of relatively lower reflectivity strategically spread over a broad wavelength range, optimizing to reduce acceleration time. This innovative approach demonstrates that a pentagonal lattice with multi-shaped periodic structures offers extra degrees of freedom, enabling the ability to tune beneficial trade-offs between reflectivity, broadband operation, and Area fraction.

A key insight is that Starshot's feasibility will hinge on balancing manufacturing costs and performance, both linked to the PhC Area fraction. A high Area fraction reduces manufacturing costs and improves yield but hurts acceleration performance. Additionally, a low Area fraction enhances acceleration but increases manufacturing complexity and costs. Neural TO navigates this optimization landscape by balancing these demands.

Integrating the cost-saving measures discussed, the total savings per sail are substantial. By reducing manufacturing costs by 9000 times and optimizing launch costs by focusing on acceleration time, we estimate significant overall budget reductions approaching 25 million euros per lightsail. This continual focus on cutting costs is essential for Starshot's success.

Future research should explore multi-objective topology optimization, incorporating structural \cite{Campbell2022}, thermal \cite{Jaffe2023, Brewer2022}, and photonic stability \cite{ilic2019self,siegel2019self,Gao2024,salary2020photonic} parameters to develop viable lightsails producible by cost-effective methods. Including realistic constraints, such as mass penalties for the lightsail’s connection to the payload, will also be crucial. 

This study underscores the potential of neural topology optimization to achieve innovative and economically viable lightsail designs, crucial for next-generation space exploration. We demonstrate that wafer-scale subwavelength-thickness reflectors can be produced in a truly scalable manner by optimizing both manufacturing costs and design. In principle, our techniques allow for the production of these low-mass, broadband reflectors at any wafer size currently available in the semiconductor industry (currently 400mm diameter). Integrating economic and performance considerations will be pivotal for the feasibility and success of ambitious projects like the Starshot Initiative. While the trajectories for such a mission are ambitious, these goals initiate a new exploration of extreme light-matter interactions, leading to advancements in photonics, structural engineering, and materials science, and opening up a new regime in these fields.

\nocite{*}
\bibliographystyle{apsrev4-1}
\bibliography{main}
\section*{Method}

\subsection*{Optimization formulation}

The neural TO algorithm is divided into four sections: convolutional neural network (CNN), post-processing, functional analysis, and calculation of the figure of merit (FOM) \cite{hoyer2019neural}. Furthermore, the optimization consists of a forward and backward step. The forward step involves feeding a randomized vector $\beta$ into the CNN, which produces the image of the optimized structure (i.e. the discretized design space).
This image is filtered, after which the performance parameters obtained in the functional analysis can be used to determine the FOM. In the backward step, the gradients with respect to the FOM are calculated for all the trainable variables of the CNN and the elements of $\beta$ so that the L-BFGS \cite{nocedal1980} optimizer can be used to update them at each new iteration. This procedure is repeated until the FOM reaches a pre-set relative tolerance or a maximum number of iterations. The performance of the neural TO approach is discussed in more detail in the supplementary information.

\begin{figure}
    \centering
    \includegraphics[width=0.75\linewidth]{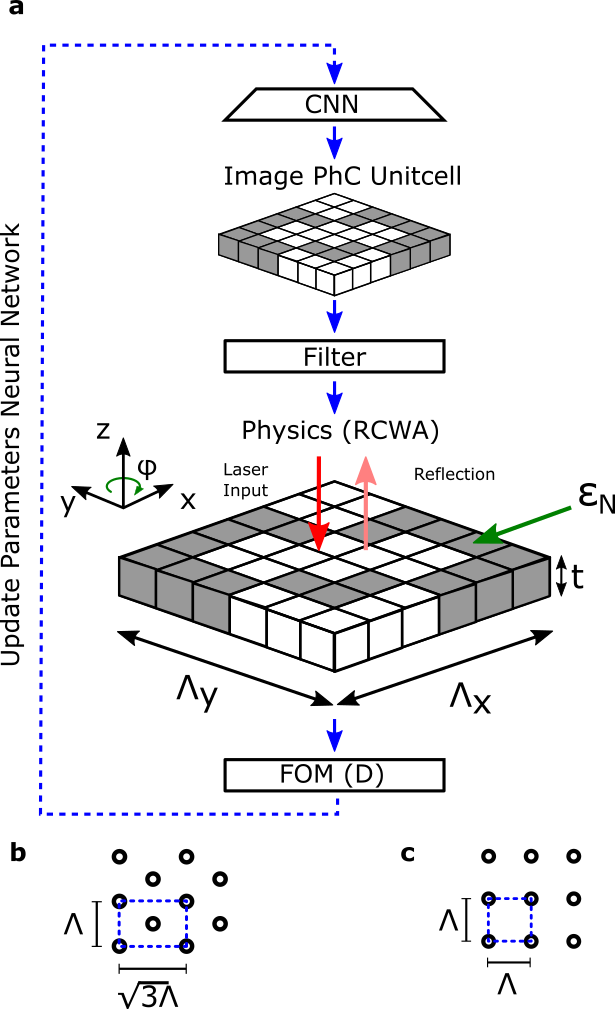}
    \caption{Schematic of lightsail optimization. \textbf{a}, The lightsail is optimized for one layer with thickness $t$. The unit cell with the period of $\Lambda$ is optimized. The discretized voxels (N) of material have an assigned dielectric constant $\epsilon_N$. The shape of the design space for a hexagonal (\textbf{b}) or square (\textbf{c}) lattice.}
    \label{fig:OptLat}
\end{figure}

For the optimization of the PhC based lightsail, only the PhC unit cell is considered. In this optimization, a two-dimensional design space is discretized into a grid of $N\times N$ (square lattice), and $N\times \sqrt{3}N$ (hexagonal lattice) pixels, and these pixels' material properties can be continuously varied between the vacuum and the required material. Furthermore, the unit cell's period $\Lambda$ (i.e. the lattice vector) and the layer thickness $t$ are used as optimization parameters. A schematic overview of the optimization and the parameters is presented in Fig. \ref{fig:OptLat}.

The functional analysis is performed with Rigorous coupled-wave analysis (RCWA) because this semi-analytical method is computationally efficient in solving scattering problems for periodic structures with layers that are invariant in the direction normal to the periodicity \cite{Jin2020}. 

Minimizing $D$ is chosen as the first objective for the design optimization. The formulation for the acceleration distance is presented in Eq. (\ref{equ:acc_dist}).

\begin{equation}
\label{equ:acc_dist}
    D=\dfrac{c^3}{2I}(\rho_l+\rho_s) \int_{0}^{\beta_f} \dfrac{h(\beta)}{R[\lambda(\beta)]} \,d\beta 
\end{equation}

In this equation, $D$ is the acceleration distance, $I$ is the intensity of the propulsion laser, $\rho_l$ and $\rho_s$ are the area densities of the lightsail and the satellite respectively, $\lambda$ is the wavelength of the propulsion laser, R is the reflection as a function of $\lambda$, and $h(\beta)=\beta/(1-\beta)^2\sqrt{1-\beta^2}$, where $\beta$ is the velocity fraction with respect to the speed of light $\beta=v/c$. Due to the Doppler red-shift of the laser, the wavelength of the laser can be written as a function of the relative speed,  $\lambda(\beta)=\lambda_0\sqrt{(1+\beta)/(1-\beta)}$. When using a 1.55 \si{\micro\meter} laser, the bandwidth at which the sail will operate is from 1.55 \si{\micro\meter} to 1.86 \si{\micro\meter}. $\rho_l$ and $R$ are the geometry-dependent parameters.

Secondly, the lightsail is optimized for $T$ \cite{Kulkarni2018}. The formulation of T is presented in Eq. \ref{equ:acc_time}.

\begin{equation}
\label{equ:acc_time}
    T=\dfrac{m_tc^3}{2IA}\int_{0}^{\beta_f} \dfrac{\gamma(\beta)^3}{R[\lambda(\beta)]} \left( \dfrac{1+\beta}{1-\beta}\right) \,d\beta 
\end{equation}

 For this equation, $m_t$ and $A$ are the total mass and area of the sail respectively, $\gamma(\beta)=1/\sqrt{1-\beta^2}$. 

The optimization aims to minimize the FOM for a SiN lightsail following the 2016 Starshot parameters. The density of SiN is set to 3100 \si{\kilo\gram/\meter\cubed} \cite{xu2024}, and its relative permittivity is 4 \cite{Steinlechner2017}. The relative permittivity of the pixels is varied between 1 (vacuum) and 4 (SiN). The thickness ($t$) and period ($\Lambda$) are constrained to $0.01~\si{\micro\meter} \leq t \leq 1~ \si{\micro\meter}$ and $0.1~ \si{\micro\meter} \leq \Lambda \leq 7.2~ \si{\micro\meter}$ respectively. The intensity $I$ of the laser beam is set to 10 \si{GW/m^2} with a wavelength $\lambda_0$ of 1.55 \si{\micro\meter}, illuminating a sail area of 10 \si{m^2}. The laser is assumed to be a linear polarised plane wave, and the sail is optimized for two orthogonal polarisation angles $\phi=0$ and $\phi=\frac{1}{2}\pi$. The initial solution of the material distribution is random, and the initial solution of the thickness and period is set to 100 nm and $\lambda_0$ respectively. The design space is divided into a 100 $\times$ 100 and 100 $\times$ 172 pixel grid for a square and hexagonal lattice respectively. 

\subsection*{Nanofabrication of the PhC}
The stringent mass requirements of the Starshot Initiative make hole-based photonic crystals (PhCs) inevitable. While multilayered and bilayer PhCs have a 100\% fill factor (i.e., no holes), making them heavier and easier to fabricate, they are unsuitable due to their excessive mass. To approach the target mass of 1 gram, we must employ single-layer PhCs with holes, achieving a fill factor of 40-70\%. This design choice, although necessary for reducing weight, introduces fragility, as stress concentrations occur in the material between holes. Consequently, fabricating centimeter-scale nanophotonic reflectors that are both lightweight and robust poses significant challenges. Effective lithography of billions of holes must be achieved rapidly, and the high-aspect-ratio single-layer PhCs must be suspended with a single, stiction-free undercut using dry chemical etching.

\begin{figure}
    \centering
    \includegraphics[width=1\linewidth]{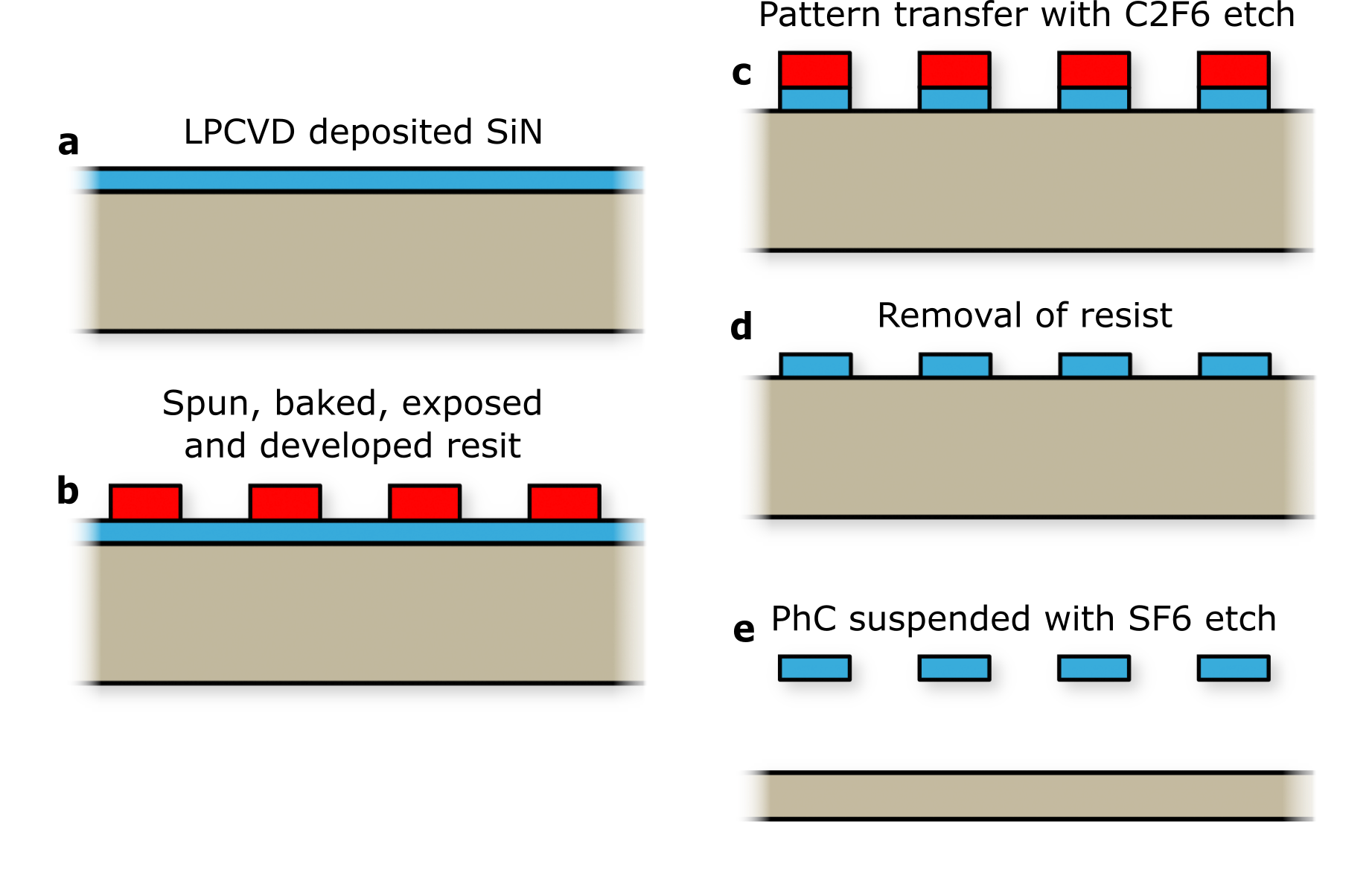}
    \caption{Schematic overview of the fabrication steps of the suspended PhC lightsail \cite{shin2022}. \textbf{a}, \ch{SiN} (blue) positioned on Si wafer with LPCVD. (\textbf{b}) Patterning of the photoresist mask (red), (\textbf{c}) Directional \ch{C2F6} plasma etch, (\textbf{d}) Resist removal, (\textbf{e}) \ch{SF6} undercut.}
    \label{fig:PHC_Fab}
\end{figure}

Fig. \ref{fig:PHC_Fab} displays the fabrication process of the suspended PhC lightsail, of which a similar process is described in the work of Shin et al. (2022) \cite{shin2022}. Initially, a 100 mm Si wafer is covered with 200 nm of silicon nitride using low-pressure chemical vapour deposition (LPCVD) to attain a pre-stress of 270 MPa. Next, the AZ ECI 3012 positive-tone resist is spin coated. Before the coating a HMDS and baking step is performed at 130 \si{\degreeCelsius} for 30 s and 60 s respectively. The resist is then spin coated at 6850 rpm to reach 1 \si{\micro\meter} thickness and soft baked at 95 \si{\degreeCelsius} for 150 s. An ASML PAS 5500/80 automatic wafer stepper is used to expose the resist with a 110 \si{mJ/cm^2} dose, operating with chrome on quartz mask to stitch $5\times5$ \si{mm^2} patterns together to a $60\times60$ \si{mm^2} PhC. The development consists of a PEB step at 115 \si{\degreeCelsius} for 150 s, followed by single puddle development with MF322 for 60 s at 3000 rpm. At last, the wafer is hard-baked at 100 \si{\degreeCelsius} for 150 s. The resist mask enables the PhC pattern to be transferred using a 60 s directional inductively coupled plasma (ICP) to etch the \ch{SiN} layer with \ch{C2F6}. Finally, a 45 s fluorine-based \ch{SF6} ICP etch is used at -120 \si{\degreeCelsius} to suspend the PhC membrane. 

\subsection*{Measurement setup for Reflectivity PhC}
The large-scale suspended PhC membrane is measured in the setup shown in Fig. \ref{fig:PHC_Meas}a to obtain the actual reflectivity spectrum and to validate the simulations. The setup consists of a tunable laser that emits a laser beam from 1530-1620 nm. This laser beam passes through an optical fibre to a circulator, followed by a polarization controller (PC). The light then goes into free space through a collimator. The light passes through a linear polarization filter to ensure it is linearly polarised. The beam size is then decreased and focused with a lens set and an objective. The reflected light follows the same path back and is diverted in the circulator to an attenuator and finally to the photodiode. The measurements are performed by first maximizing the signal retrieved from a silver mirror at 1580 nm, after which its reflectivity spectrum is measured, and this is repeated for the PhC membranes. The actual reflectivity could be obtained by normalising the measurement of the PhC membrane with the known reflectivity of the silver mirror. Fig. \ref{fig:PHC_Meas}b shows the 100 mm diameter wafer within the measurement setup.

\subsection*{Simulation of measurement}
The PhC unit cell of the lightsail is simulated as a \ch{SiN} membrane surrounded by a vacuum in the lightsail TO. However, the fabricated design is a suspended \ch{SiN} membrane attached to a Si wafer. Therefore, to match the measurements with the simulations, the full system must be considered as presented in Fig. \ref{fig:PHC_Meas}d. This figure shows the vacuum gap between the \ch{SiN} membrane and the Si wafer, which is around 4 \si{\mu m} and can be deducted from the undercut at the edge of the membrane through optical microscopy. Additionally, a layer around 100 nm with a refractive index of $(\epsilon_{Si}+\epsilon_{vac})/2$ \cite{Fujiwara2018} was added on top of the Si wafer, representing the roughness resulting from the undercut. This resulted in a qualitative match of the measurement, yet the total reflectivity did not match. This mismatch is because the setup is designed only to measure the normally reflected, zero-order diffracted light. However, the total reflectivity is used in the lightsail optimization to obtain the total momentum transfer in the normal direction of the sail. Consequently, after fitting the height of the gab, the roughness layer and when only considering the zero-order reflected light, the simulation is in good agreement with the measurements.

\section*{Acknowledgements}
 We want to thank Yufan Li, Peter Steeneken, Megha Khokhar, Mark Kalsbeek, Juan Lizarraga Lallana and Ata Keskeklar for stimulating discussions. Additionally, we want to thank R. Tufan Erdogan for his support with our measurements. Funded/Co-funded by the European Union (ERC, EARS, 101042855). Views and opinions expressed are however those of the author(s) only and do not necessarily reflect those of the European Union or the European Research Council. Neither the European Union nor the granting authority can be held responsible for them. R.N. and M.A.B would like to acknowledge support from the Limitless Space Institute’s I$^2$ Grant.

\clearpage



\section*{Supplementary material}

# Supplementary Information on "Pentagonal Photonic Crystal Mirrors: Scalable Lightsails with Enhanced Acceleration via Neural Topology Optimization"

July 4, 2024

L. Norder, S. Yin, M. J. de Jong, F. Stallone, H. Aydogmus, P.M. Sberna, M. A. Bessa, R. A. Norte,

**Section A - Design analysis**

24 evaluation points are used in calculating the $D$ in the optimization to reduce the computational cost, resulting in an error to the actual $D$. Therefore, the three designs from Fig. S1 closest to the 500 nm mean feature size (MFS) objective are studied more thoroughly to choose a final design for this study. Firstly, the full reflectivity spectrum of the four designs is calculated for 300 wavelengths and presented in Fig. S1b. These spectra are then used to integrate the equation of motion directly [4], resulting in the velocity and the travelled distance of the lightsails following the Starshot parameters (Fig. S1d,e), giving insight into how the reflectivity spectrum translates to the $D$. It stands out that for designs with similar $D$ the actual acceleration time can vary significantly.

Secondly, the design polarisation dependency is studied by plotting the $D$ of the sail for a single plane wave normal to the sail surface by varying from $\phi = 0$ to $\phi = \pi$, because during the launch the lightsail will only be illuminated by a single linear polarised plane wave. Fig. S1c shows that the three designs perform differently. Whereas the $D$ varies negligibly for the hexagonal design, the square design loses all its performance between $\phi = 0.5\pi$ and $\phi = \pi$. Therefore, the feasibility of PhC design depends on the alignment accuracy or the envisioned operating state (e.g. spinning lightsails).

It should be noted that the final acceleration time and distance presented in Fig. S1e are probably not feasible for a physical lightsail, as it is not considering the connection of the sail to the payload. Although this study has shown some practical limits to the lightsail missions, it would be valuable to study the limits of

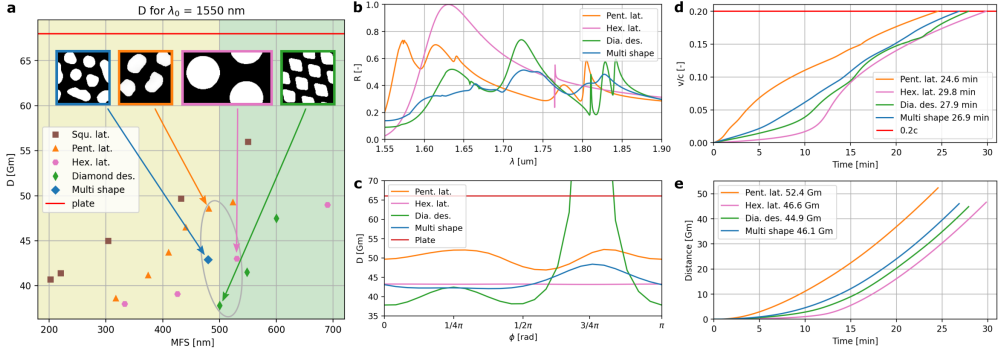

Fig. S 1: Performance evaluation of the obtained designs. **a**, Acceleration distance $D$ of obtained designs for various MFS. **b**, Reflectivity spectrum. **c**, $D$ for one incident plane wave with changing polarisation angle. The velocity (**d**) and the travelled distance (**e**) of a PhC lightsail over its acceleration time.

| Source | Spectrum | $\lambda$[nm] | MLW [nm] | system cost [$M] |
|---|---|---|---|---|
| Hg i-line | UV | 365 | 350 | 4-6 |
| KrF | DUV | 248 | 150 | 7-11 |
| ArF | EUV | 193 | 19 | 25-110 |

Table 1: Minimal line width (MLW) for optical lithography using different light sources [5, 1]

the different subsystems (e.g. laser array and lightsail) to improve the optimization and feasible design generation. Moreover, Fig. S1c,d shows that optimizing for $D$ alone is insufficient to capture the challenging mission criterion.

To summarise, optimising a lightsail for $D$ needs to consider designs with varying MFS, polarisation dependence, and acceleration time. These three parameters can highly influence the mission's success by determining the fabrication and the launch costs.

### Section B - System costs optical lithography

Multiple optical lithography systems with different light sources and, therefore, different minimum line widths are presented in TABLE 1. However, the MFS of this lithography method will exceed the minimum line width (MLW) due to the diffraction limit of the light. Therefore, sharp features like curvatures may not be captured during production, resulting in an MFS of 500 nm. For this study, i-line photolithography was selected based on its cost-effectiveness, availability, and established processing protocols.

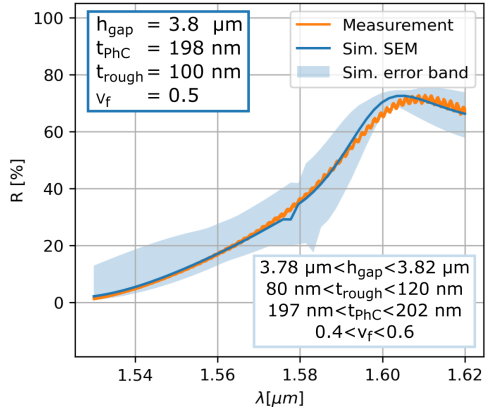

Fig. S 2: Reflectivity measurement from the $60 \times 60$ mm$^2$ suspended photonic crystal membrane together with the RCWA simulation of the reflectivity and related error band. $h_{gap}$, the height of vacuum gap between the membrane and the Si substrate; $t_{PhC}$, the thickness of the membrane; $t_{rough}$, the thickness of the roughness layer in the simulation; $v_f$, the volume fraction of the roughness layer which determines the relative permittivity of the layer by $\epsilon_{rough} = \epsilon_{Si} v_f + \epsilon_{SiN}(1-v_f)$ [2].

## Section C - Measurement simulations

To accurately represent the measurement, a new simulation is performed, in which not only the membrane in a vacuum but also the Si wafer and the gap between the wafer and the membrane are simulated. Additionally, a small layer with relative permittivity between air and Si is added to simulate the roughness of the Si substrate due to the fabrication. Fig. S2 shows a representative measurement that is fitted to the model by varying the height of the SiN layer, the gap between the Si and SiN, and the layer representing the rough Si surface.

## Section D - Mass constraint

In the context of the Starshot initiative, the total mass of the lightsail has been suggested to be 1 gram. The design we found, as reported in the main text and illustrated in Fig. 3, has an approximate mass of 3 grams for a 10 $m^2$ lightsail using the selected material. We have investigated the feasibility of creating a lightsail with less mass by adjusting the area fraction and the sail's thickness, ensuring a particular total mass value is satisfied. Fig. S3 illustrates the outcome of the optimization for different mass constraints when considering different sail thicknesses. We found that the designs obtained for a mass constraint of 1 gram are not viable for manufacturing (labelled case 1 in the figure) because they have freely suspended masses. In fact, the main factor contributing to a large acceleration distance is the low mass of these sails, rather than an increase in reflectivity. We also highlight that the best designs

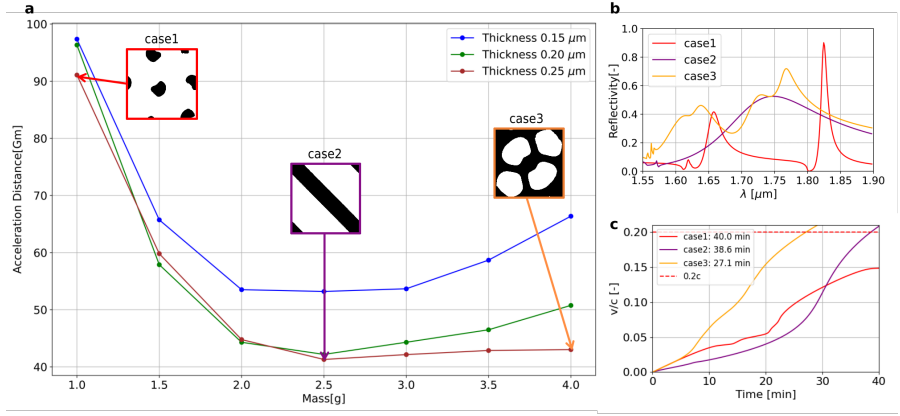

Fig. S 3: Best design for different mass constraints. **a**. Limiting the lightsail to 1 gram precludes manufacturability, as shown in Case 1. Case 2 yields the shortest acceleration distance yet suffers from significantly reduced reflectivity compared to Case 3, which exhibits a marginally greater acceleration distance. **b**. Reflectivity spectrum for these three cases. Case 3 has a relatively lower acceleration distance but the highest reflectivity. **c**. Case 3 has the lowest acceleration time because of the high reflectivity.

we found for a lightsail with 2.5 grams are still not viable for manufacturing, as they would correspond to parallel strings of material – case 2 in the figure. Furthermore, we show in part Fig. S3b,c that the lower reflectivity across the laser wavelength spectrum leads to a higher acceleration time, which would require more exposure to the laser light. Therefore, the pentagonal crystal structure found for larger mass values is more interesting, as it has lower acceleration time and it is viable for manufacturing.

### Section E - Neural topology optimization

In this study, we investigated for the first time the use of neural topology optimization in Photonics by considering a convolutional neural network that reparameterizes the lightsail design on the fly (without training). This recent method has been evaluated for structural optimization tasks, but we believed that it would be more advantageous in the context of lightsail design because conventional topology optimization requires a relaxation factor $Q$ to ensure a smooth optimization process [3]. The conventional strategy utilizes a pixel-based model and the Method of Moving Asymptotes (MMA) as the optimizer. We implemented this strategy using the NLopt library, referring to it as "Pixel model", and compared it with the neural TO method.

In the absence of a relaxation factor the simulation is closer to the real physical problem, but the conventional TO method does not converge to a good solution.

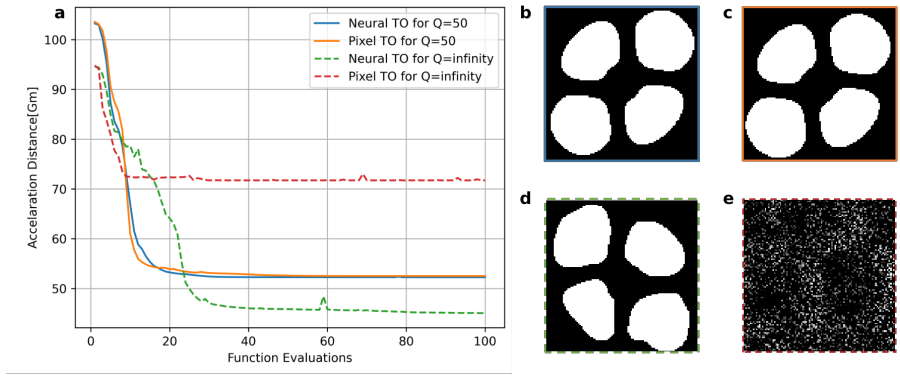

Fig. S 4: Comparative performance of the "Pixel" and neural TO models on a fabricated unit cell with a periodicity of 3.0 $\mu m$ and thickness of 200 $nm$. **a** Loss curves for the two models, with and without the relaxation factor. **b** and **c** show the similarity in design and objective values achieved by the "Pixel" TO model and neural TO model, respectively, when a relaxation factor is applied. Without the relaxation factor, as shown in panels **d** and **e**, the "Pixel" model does not converge to a good solution, whereas the neural TO model successfully converges.

.

Even when considering more than 100 iterations, there is virtually no change in the design for the conventional strategy. In contrast, our findings demonstrate that the neural TO method finds a similar solution (pentagonal crystal) without the need for this relaxation factor, while predicting a lower (better) objective value. The reparameterization technique allows to handle objective landscapes that are non trivial and non-convex by over-parameterizing the problem and avoids getting stuck in local optima. We found this method to be robust and we did not have to invest significant time in hyperparameter optimization. We believe that neural TO will open new avenues in design for problems that may involve even more complex objective landscapes in the future.



\end{document}